\documentclass{article}
\usepackage{ijcai26}

\usepackage{times}
\usepackage{soul}
\usepackage{url}
\usepackage[hidelinks]{hyperref}
\usepackage[utf8]{inputenc}
\usepackage[small]{caption}
\usepackage{graphicx}
\usepackage{amsmath}
\usepackage{amsthm}
\usepackage{booktabs}
\usepackage{algorithm}
\usepackage{algorithmic}
\usepackage[switch]{lineno}
\usepackage[misc]{ifsym}
\usepackage{verbatim}

\title{From Surface Learning to Deep Understanding: \\A Grounded AI Tutoring System for Moodle
}

\author{
Anna Ostrowska, Michał Kukla, Gabriela Majstrak, Jan Opala, Sebastian Pergała, Jan Skwarek, Anna Wróblewska \Letter
\affiliations
Faculty of Mathematics and Information Sciences, Warsaw University of Technology, Warsaw, Poland
\emails
\Letter~anna.wroblewska1@pw.edu.pl
}

\begin{document}

\maketitle

\begin{abstract}
    
This demo paper describes the development of the AI Teaching \& Learning Assistant, a modular Moodle plugin that leverages Retrieval-Augmented Generation (RAG) to deliver high-quality, hallucination-free education. The system employs a dual-centric design, providing students with interactive, Socratic-based tutoring and educators with a "Human-in-the-loop" workspace for supervised content generation. By grounding Large Language Model (LLM) responses in teacher-provided materials, the assistant addresses the risks of misinformation while encouraging deep conceptual mastery. Evaluation via the Ragas (LLM-as-a-Judge) framework and a preliminary user study confirms its effectiveness, achieving faithfulness scores up to 0.97 and a 4.00/5.00 recommendation rate.


 \textbf{Demo Video:} \url{https://tinyurl.com/mrytc2dr} \\
\end{abstract}

\section{Introduction}

Modern e-learning platforms like Moodle have revolutionized access to education, but they have also introduced information overload \cite{algahtani2011evaluating}. Students often face a massive volume of files and deadlines, leading to "surface learning", a strategy where learners focus on meeting immediate requirements rather than achieving a profound understanding of the subject \cite{motlagh2023impact}. Concurrently, while general AI tools have become popular, they present significant risks: they are prone to hallucinations (generating factually incorrect but plausible-sounding information), they lack institutional integration, and high-quality versions are often expensive, creating an accessibility gap \cite{ahmed2024digital}.


The primary objective of this project is to transform AI from a passive information provider into an active, supervised learning partner. The system aims to:
\textbf{minimize misinformation} by grounding all AI outputs in verified course materials; \textbf{reduce teacher workload} through the semi-automation of "tedious" tasks like quiz and study guide preparation; foster \textbf{critical thinking} by using pedagogical frameworks like the Socratic method and Bloom's Taxonomy; ensure \textbf{data privacy} through a modular architecture that isolates data between different courses.
\noindent The application follows a dual-centric approach: 
\textbf{Teacher-Centric Use Cases}: The system facilitates the didactic process by automating the preparation of study materials and tests. It maintains a "Human-in-the-loop" workflow, where teachers act as essential validators, reviewing and approving AI-generated content before it is released to students. 
\textbf{Student-Centric Use Cases}: For learners, the assistant provides an interactive chatbot and a personalized quiz module. These features are designed to move students away from passive "surface learning" toward a deep understanding of concepts and long-term retention. \\
The system is designed to be cost-effective and easy to set up; it employs a modular design with a FastAPI backend, a Next.js frontend, and a PHP-based Moodle plugin, supporting both local deployment and offloading external computation to ensure scalability.

\section{Related Work \& Market Analysis}

The \textbf{evolution of artificial intelligence in education} has shifted from early rule-based and pattern-matching systems, such as ELIZA \cite{perez2020rediscovering}, which lacked conversational memory and produced deterministic responses, to more advanced systems that use Latent Semantic Analysis (LSA) to understand student responses through semantic similarity and adapt teaching strategies \cite{adamopoulou2020overview}. Modern educational chatbots have further advanced this field by combining conversational modules with information retrieval components, enabling dynamic updates to knowledge bases with new academic materials \cite{berry2023limits}. In current research, a primary methodological distinction is made between Fine-Tuning, which modifies a model's internal weights at a high computational cost, and Retrieval-Augmented Generation (RAG), which grounds large language model responses in external, non-parametric memory \cite{lewis2020rag}. RAG is increasingly preferred in academic contexts because it enables easy updates to course content and significantly mitigates the risk of hallucinations by providing verifiable citations to teacher-provided sources.

To maintain \textbf{high pedagogical standards}, modern systems integrate established strategies such as the Socratic method, which encourages active thinking by guiding students toward discovery rather than providing direct answers, and Bloom's Taxonomy, which structures assessments across different cognitive levels \cite{guldal2025can,gadhiya2025chatbot}. The evaluation of these RAG-based educational tools has also moved beyond traditional NLP metrics toward specialized frameworks such as Ragas, which assess performance across dimensions of faithfulness, answer relevance, and context recall \cite{nguyen2022design,dan2025applications,ji2023survey}. 

Currently, \textbf{the market} for AI educational tools includes specialized platforms such as StudyFetch \cite{studyfetch2026}, NotebookLM \cite{notebooklm}, and Quizbot \cite{quizletQChat}, which generate learning resources, such as flashcards and quizzes, from uploaded documents. Additionally, major providers offer institutional solutions such as ChatGPT Edu \cite{openaiChatGPTEdu}, Gemini for Education \cite{googleGeminiEducation}, and Microsoft Copilot for education \cite{microsoftCopilotEducation}, designed to increase academic productivity and summarize texts.

Despite these advancements, a significant \textbf{"universal integration gap"} persists, as most existing prototypes operate as standalone services that require manual file uploads and lack native integration with established Learning Management Systems (LMS) such as Moodle \cite{hunt2012moodle,moodle_docs}. Furthermore, there is a critical shortage of open-source systems that support a "Human-in-the-loop" workflow, in which teachers act as essential validators, reviewing and approving AI-generated content before it is released to students. The research presented here addresses these shortcomings by offering a modular, cost-effective solution that transforms AI from a passive information provider into an active, teacher-supervised partner in the learning process.

\section{System Architecture}
\begin{figure*}
    \centering
    \includegraphics[width=0.65\linewidth]{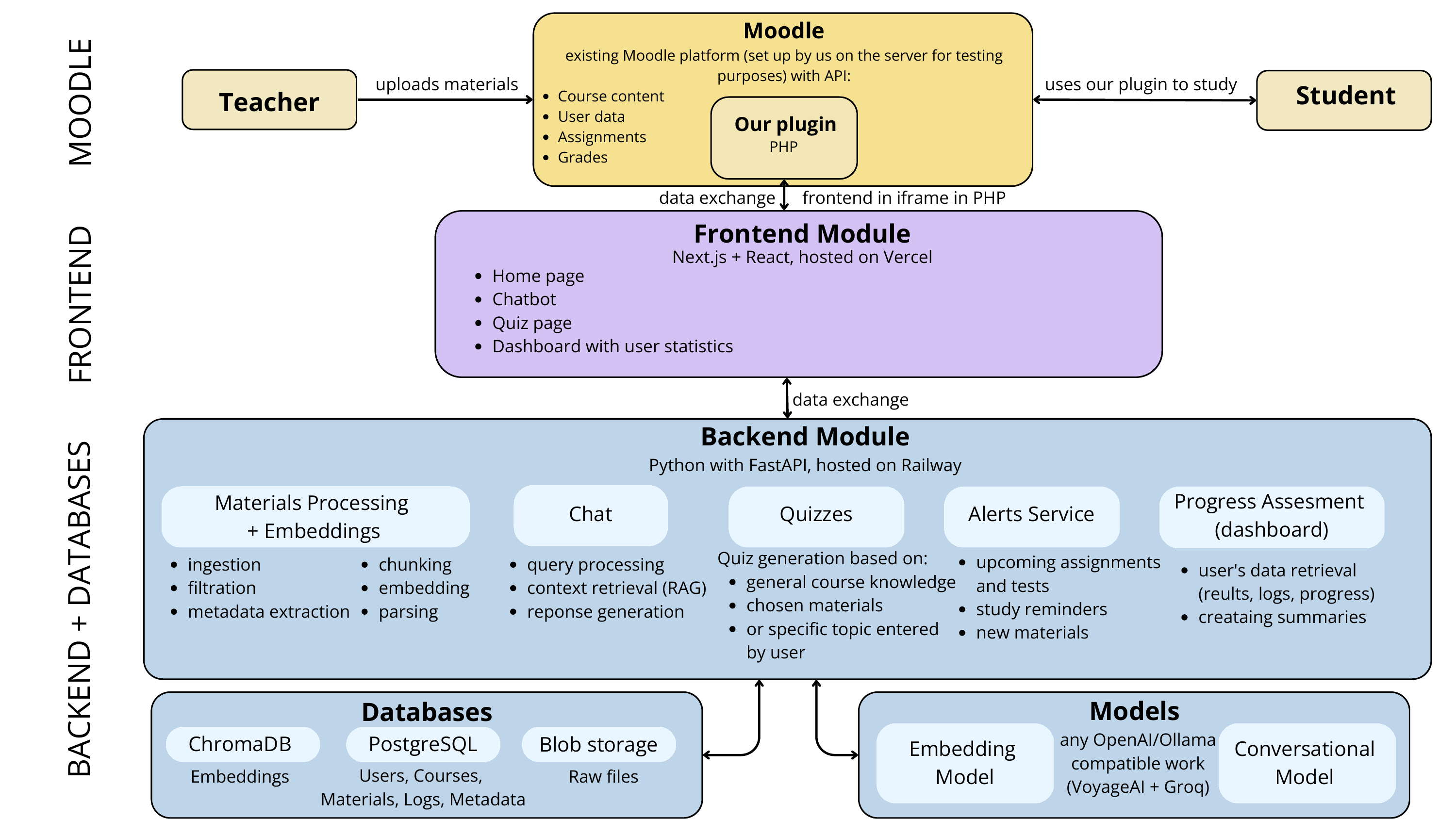}
    \caption{Our system architecture: Presentation Layer  --Moodle Plugin and Frontend Module, Application Layer -- Backend Module, Data Layer - Databases and Models}
    \label{fig:placeholder}
\end{figure*}
The system employs a three-layer -- \textbf{Presentation, Application, and Data} -- modular architecture designed for scalability and reproducible deployment. 
\paragraph{Presentation Layer (Moodle Plugin).} The primary interface is built around a native PHP-based Moodle Activity Module plugin \cite{moodle_plugin_types}, which serves as the secure entry point for all users. For teachers, this layer serves as an administrative workspace with a "Human-in-the-loop" editor, where AI-generated materials and quizzes are held in "unreviewed" status until they are manually validated and approved. For students, the layer utilizes a Next.js and React frontend embedded within a Moodle IFrame to provide a responsive environment for interactive tutoring and self-assessment. To ensure academic content remains legible, the presentation layer integrates libraries like KaTeX and Marked.js to accurately render complex LaTeX mathematical formulas and Markdown-formatted text.
\paragraph{Application Layer (FastAPI Backend).} A Python-based FastAPI backend serves as the API Gateway, coordinating the RAG pipeline and AI services. This layer handles: \textbf{parsing} -- extracting text from PDF, DOCX, and PPTX files; \textbf{intent recognition} -- a lightweight all-MiniLM-L6-v2 sentence-transformer model automatically classifies user prompts into "Explanation," "Test Generation," or "Material Generation" modes; \textbf{asynchronicity} -- all pipelines run asynchronously to prevent users from blocking one another during long-running tasks like document indexing.
\paragraph{Data Layer (Hybrid Storage).} To manage diverse data types and roles, a hybrid strategy is used: \textbf{PostgreSQL} stores structured relational data (course info, user metadata, conversation logs); \textbf{ChromaDB} -- an AI-native vector database that stores high-dimensional embeddings for rapid semantic search; \textbf{MinIO} -- an S3-compatible blob storage for large raw files, keeping the relational database performant.

\paragraph{Deployment and Orchestration.} The entire integrated environment is containerized using Docker, allowing the various services—including Moodle, FastAPI, the databases, and storage servers -- to run in isolated, reproducible environments. 

\section{Interaction Design \& Interface}
The interaction design and interface of the AI Learning Assistant provide a seamless, integrated experience within the \textbf{Moodle Learning Management System} (LMS) that supports both teaching and learning workflows, which can support each other interchangeably (also see~\cite{wroblewska2025evaluating}). 

\paragraph{Student-Centric Interface.} Students access the system via a Home Page hub featuring three workflows.
\textbf{Conversational Chatbot} features three modes—Quick Answer (fact-based), Deep Understanding (Socratic guidance), and Exam Coach (Chain-of-Thought reasoning). Additionally, \textbf{Source Reader} allows students to expand AI responses to see exact page numbers and document fragments.
\textbf{Quiz Center} generates personalized tests based on the entire course or specific topics. After submission, students receive detailed explanations linked to course materials. \textbf{Learning Dashboard} visualizes progress through non-linear coverage metrics, showing which materials are "Completed" or "In Progress" based on interaction frequency.


\paragraph{Teacher-Centric Workflow.} Educators use the Teaching Assistant tab to manage course context.
\textbf{Human-in-the-loop Editing}: ensures that all AI-generated tests are initially "Unreviewed". Teachers use a built-in Markdown editor to refine questions.
\textbf{Moodle Integration} enables validated assessments to be exported in the Moodle XML format, allowing them to be imported directly into the institutional Question Bank.
To further personalize the environment, a Settings panel lets users adjust the visual atmosphere 
ensuring the tool remains accessible and engaging across devices.

\section{Experimental Evaluation}
\paragraph{Technical Optimization (LLM-as-a-Judge).} The evaluation of the system using the RAGAS (Retrieval-Augmented Generation Assessment) framework provided quantitative insights into the performance of the retrieval and generation pipelines across different academic disciplines and interaction modes \cite{ragas_paper,ragas_docs}. 
Our experiments revealed that the optimal system configuration differs significantly depending on the subject matter:
for STEM subjects: The optimal chunk size was 512 characters, and the best performance was reached with a model temperature of 0.3 were optimal for technical formulas, achieving a RAGAS faithfulness of 0.97; for humanities: larger chunks (1000 chars) and a lower temperature (0.1) were required to capture interpreted meanings, achieving a context recall of 1.00, and an average faithfulness of 0.98.
Across both disciplines, Gemini Flash 2.0 was chosen for production due to its superior faithfulness and speed-to-cost ratio compared to GPT-4o-mini and Llama 3.1. For retrieval settings, the "sweet spot" for context was found to be Top-K = 10; increasing this to 15 did not improve recall but significantly worsened context precision due to the "distraction issue" caused by irrelevant information.

The Ragas evaluation of the three specialized interaction modes confirmed that they successfully achieved pedagogical goals:
\textbf{Quick Mode} achieved the highest faithfulness (0.92), prioritizing direct, factual answers grounded strictly in the source materials; \textbf{Deep Understanding} intentionally recorded low answer relevancy because its Socratic design focuses on asking guided questions rather than providing answers, yet it maintained near-perfect context recall (0.98); \textbf{Exam Coach} successfully balanced structured support with accuracy, yielding the highest answer relevancy (0.82) through the use of Chain-of-Thought reasoning.

\paragraph{Human Feedback Analysis.}
The experiments with human feedback were conducted to evaluate the system's usability, effectiveness, and pedagogical value from both students' and educators' perspectives.\footnote{We have ethical approval from our university's Ethics Board to conduct experiments with users and to administer the survey.} These studies utilized a Likert scale (1–5) and open-ended qualitative questions to gather detailed insights into the user experience.
The evaluation involved 18 participants in the final study, of whom the majority were students (83.3\%) and instructors (11.1\%) from our university. Participants were tasked with interacting with the chatbot, generating and solving quizzes based on real academic materials (such as NLP and Machine Learning course notes), and navigating the learning dashboard to track their progress. 
The system was highly successful in its primary goal of providing grounded, reliable information. Participants consistently rated the resistance to hallucinations as high (Mean = 4.44/5), confirming that RAG was effectively limited to teacher-provided sources. The overall recommendation score for the system was 4.00 out of 5, with the highest individual ratings given to the ease of starting a conversation (4.39/5) and the relevance and accuracy of responses (4.06/5). Furthermore, 14 out of 18 users identified the chat interface and automatic test generation as the most useful features of the assistant and the source reader for academic transparency.  Teachers valued the ability to monitor student logs and maintain control over content generation.


\section{Summary and Impact}


Our AI Tutor bridges the gap between conversational AI and institutional Learning Management Systems. 
By providing an open-source, cost-effective solution, the assistant eliminates financial barriers to high-quality AI education while maintaining strict data isolation and privacy between courses. For educators, the system drastically reduces the time-intensive burden of manual content creation without sacrificing control, as every AI-generated quiz or study guide must be manually approved before students receive it. By strictly grounding AI in teacher-verified documents, the system moves students away from "surface learning" toward a grounded mastery of their academic subjects.
Ultimately, this project transforms AI from a passive information provider into an active, supervised partner in the learning process. 
Our AI tutor can be expanded in many directions, e.g., deepening research on grounded RAGs, fully free from hallucination; understanding and generating multimodal content; and making observations on users' learning habits based on interaction log data collected on an ongoing basis.

\section*{Acknowledgements}
This work was supported by the European Union under the Horizon Europe grant "Overcoming Multilevel Information Overload" (OMINO, \url{https://ominoproject.eu}, grant no. 101086321) and by the Polish Ministry of Education and Science within the framework of the program titled International Projects Co-Financed.\footnote{However, the views and opinions expressed are those of the authors only and do not necessarily reflect those of the European Union or the European Research Executive Agency. Neither the European Union nor the European Research Executive Agency can be held responsible for them.} We also thank Dr. Daniel Dan (Modul University Vienna) for providing an initial introduction to the task of developing software integrated with Moodle LMS, and the survey participants for taking the time to test our system and providing valuable feedback.

\section*{Contribution Statement}

\textbf{Anna Ostrowska} \textbf{14\%}, Developed the Moodle Plugin architecture and Chatbot mode logic, \emph{CRediT: Methodology, Project administration, Software, Visualization, Writing – original draft}; \\
\textbf{Michał Kukla}  \textbf{13\%}, Architected the PostgreSQL relational database and intent recognition, \emph{CRediT: Methodology, Project administration, Visualization, Writing – original draft, Software}; \\
\textbf{Gabriela Majstrak}  \textbf{13\%}, Led the system-wide RAG optimization and LLM comparative testing, \emph{CRediT: Formal analysis, Investigation, Methodology, Software, Visualization, Writing – original draft}; \\
\textbf{Jan Opala} \textbf{13\%}, Managed global AWS/Moodle infrastructure and led RAGAS/IRT validation, \emph{CRediT~\cite{credit}: Formal analysis, Investigation, Methodology, Resources, Visualization, Software, Writing – original draft, Data curation}; \\
\textbf{Sebastian Pergała}  \textbf{14\%}, Engineered the backend ingestion pipeline and retrieval logic, \emph{CRediT: Validation, Visualization, Writing – original draft,  Software}; \\
\textbf{Jan Skwarek}  \textbf{13\%}, Handled FastAPI security, authorization, and backend initialization, \emph{CRediT: Data curation, Validation, Visualization, Writing – original draft, Software}; \\
\textbf{Anna Wróblewska}  \textbf{20\%} Scientific Lead, Strategy, Funding, and Oversight, \emph{CRediT: Conceptualization, Funding acquisition, Investigation, Methodology, Project administration, Resources, Supervision, Validation, Writing – original draft, Writing – review \& editing}.   \\

\noindent All authors have read and agreed to the published version of the manuscript.

\bibliographystyle{named}
\bibliography{ijcai26}
\end{document}